\documentstyle[11pt]{article}
\newcommand{\be}{\begin{equation}}
\newcommand{\ee}{\end{equation}}
\newcommand{\ba}{\begin{eqnarray}}
\newcommand{\ea}{\end{eqnarray}}
\newcommand{\bb}{\begin{thebibliography}}
\newcommand{\eb}{\end{thebibliography}}
\newcommand{\ci}[1]{\cite{#1}}
\newcommand{\bi}[1]{\bibitem{#1}}
\newcommand{\lab}[1]{\label{#1}}

\begin{document}

\begin{center}
{\large \bf{Spin effects in high energy diffractive
reactions.}}
\\
S.V.Goloskokov \\
  Bogoliubov Laboratory of Theoretical  Physics,
  Joint Institute for\\  Nuclear Research,
  Dubna 141980, Moscow region, Russia.

\vspace{.3cm}
\begin{abstract}
It is shown that single  and
double spin asymmetries in  polarized diffractive
$Q \bar Q$ production depend strongly on the spin structure of the
quark-pomeron vertex. They can be studied in future spin experiments at
HERA.
\end{abstract}

\end{center}

Experimental observations of the diffractive production of high $p_t$ jets
in hadron-hadron and lepton-proton reactions \ci{ua8,h1}
have stimulated the study of pomeron properties.
Diffractive reactions with high $p_t$ jets can be interpreted in terms of the
partonic structure of a pomeron \ci{ing}. It has been shown that the observed
effects can be predominated by the quark structure of a pomeron
\ci{lanj,coll}.

The pomeron  is a vacuum $t$-channel exchange that contributes to high-energy
diffractive reactions. The  nonperturbative two-gluon
exchange model  \ci{low,la-na} and the BFKL model \ci{bfkl} of a pomeron
lead to the mainly imaginary scattering amplitude
that can be written as a product of pomeron vertices
$V^{hh I\hspace{-1.1mm}P}_\mu$ and some function $I\hspace{-1.6mm}P$ determined
by the pomeron. As a result, the pomeron contribution to
the quark-proton amplitude looks as follows
\be
T(s,t)= i I\hspace{-1.6mm}P(s,t)\; V_{qqI\hspace{-1.1mm}P}^{\mu} \otimes
V^{ppI\hspace{-1.1mm}P}_{\mu}.    \lab{tpom}
\ee

In models \ci{la-na,bfkl} the pomeron couplings have a simple matrix
structure:
\be
V^{\mu}_{hh I\hspace{-1.1mm}P} =\beta_{hh I\hspace{-1.1mm}P}\; \gamma^{\mu},
\lab{pmu}
\ee
which reflects the well-known approximation that the spinless quark-pomeron
coupling is like a $C=+1$ isoscalar photon vertex. In this case the spin-flip
effects are very small.

However, the spin structure of quark-pomeron coupling may be not so simple.
 The perturbative calculations \ci{gol-pl} show the quark-pomeron vertex
complicated in form :
\be
V_{qqI\hspace{-1.1mm}P}^{\mu}(k,r)=\gamma_{\mu} u_0+2 m k_{\mu} u_1 +
2 k_{\mu}
/ \hspace{-2.3mm} k u_2 + i u_3 \epsilon^{\mu\alpha\beta\rho}
k_\alpha r_\beta \gamma_\rho \gamma_5+im u_4
\sigma^{\mu\alpha} r_\alpha.    \lab{ver}
\ee
Here $k$ is a quark momentum, $r$ is a momentum transfer.
The structure of the quark-pomeron vertex function (\ref{ver}) is
drastically different from the standard pomeron coupling (\ref{pmu}).
Really, the terms
$u_1(r)-u_4(r)$ lead to the spin-flip in the quark-pomeron vertex in contrast
to the term proportional to $u_0(r)$.  The functions
$u_1(r) \div u_4(r)$ at large $r^2$ are not very small \ci{gol4}. Note that
the phenomenological  vertex $V_{qqI\hspace{-1.1mm}P}^{\mu}$  with $u_0$ and
$u_1$ terms has been proposed in \ci{klen}.

The proton-pomeron vertex has been found in some models (see \ci{golpp} e.g.):
\be
V_{ppI\hspace{-1.1mm}P}^{\mu}(p,r)=m p_{\mu} A(r)+ \gamma_{\mu} B(r),
\lab{prver}
\ee
which is similar in form to that from Ref. \ci{klen}.
The quantity $A$ in (\ref{prver}) determines the transverse polarization and
$B$ contributes to the longitudinal asymmetry.

Thus, the pomeron vertices have a complicated spin structure. This  should
modify different spin
asymmetries in high energy diffractive reactions
that can be measured, for example, in future spin experiments at HERA.

We would like to show that one of the simplest way to test the quark-pomeron
vertex is to study the $Q \bar Q$ production in diffractive reactions.

 In this report we shall discuss the longitudinal double spin asymmetries
in  polarized $p\uparrow p\uparrow \to p+Q \bar Q+X$
and  $l\uparrow p\uparrow \to l+p+Q \bar Q$ diffractive reactions,
single transverse spin asymmetry in similar reactions.
Such experiments will be possible in the future HERA-N project and at HERA
with a polarized proton beam.

Let us analyse the $Q \bar Q$ production in diffractive $pp$ scattering.
We shall investigate graphs where the pomeron with a nonzero momentum transfer
interacts with one quark in the loop. The integration over all the $Q \bar Q$
phase space will be performed.
It has been shown in our previous estimations \ci{golasy} that
$A_{ll}$ asymmetry in this case can reach $10 \div 12 \%$.

The standard kinematical variables look as follows
\be
s=(p_i+p)^2,\; t=r^2=(p-p')^2,\;x_p=\frac{p_i(p-p')}{p_i p}.  \lab{varpp}
\ee
Here $p_i$ and $p$ are initial proton
momenta, $p'$ is a momentum of a recoil hadron, $r$ is a momentum transfer at
the pomeron vertex and $x_p$ is a part of the momentum $p$ carried off by the
pomeron.
The investigated process is important at small $x_p$ that lead to a small
invariant mass in the $Q\bar Q$ system  $M_x^2 \sim \bar x_p s$.
The diagram with a triple pomeron vertex must be considered for large
$M_x^2$. Then we shall see more than two high $p_t$ jets.

We shall study the region where $|t|$ is a few $GeV^2$ and
$x_p \sim 0.1 \div 0.2$ . For this $|t|$ the perturbative QCD can be used
for calculating the spin structure of the quark-pomeron vertex.

We shall investigate the longitudinal double spin asymmetry determined by
the relation
\be
A_{ll}=
\frac{\Delta \sigma}{\sigma}=\frac{
\sigma(^{\rightarrow} _{\Leftarrow})-\sigma(^{\rightarrow} _{\Rightarrow})}
{\sigma(^{\rightarrow} _{\Rightarrow})+\sigma(^{\rightarrow} _{\Leftarrow})}.
\lab{asydef}
\ee
For the spin-average and longitudinal polarization of the proton beam
the $B$ term in (\ref{prver}) is predominant. As a result, the longitudinal
double spin asymmetry does not depend on the pomeron-proton vertex structure.

We find that $\sigma \propto 1/x_p^2$ at small $x_p$. This behaviour is
associated usually with the pomeron flux factor for
$\alpha_{I\hspace{-1.1mm}P}(0)=1$. However, $\Delta\sigma$ is proportional to
$\epsilon^{\mu\nu\alpha\beta}r_\beta...\propto x_p p$.
Thus, additional $x_p$
appears and we find that $\Delta\sigma \propto 1/x_p$ at small $x_p$.

In calculations of the corresponding integrals, the off-mass-shell behaviour
of the pomeron structure functions
$u_i$ has been considered. The
simple form of the $u_0(r)$  function
$$   u_0(r)=\frac{\mu_0^2}{\mu_0^2+|t|} ,\;\;\;r^2=|t|,$$
 was used with $\mu_0 \sim 1Gev$ introduced in  \ci{don-la}. The functions
$u_1(r) \div u_4(r)$ at $|t|>1 GeV^2$ were calculated in perturbative QCD
\ci{gol4}.

The main contribution to $\Delta \sigma$ is proportional to the first moment
of $ \Delta g$
\be
 \Delta g =\int_{0}^{1} dy \Delta g(y), \lab{i3}
\ee
which is unknown up to now.
In explanation of the proton spin \ci{efr} a large magnitude of
 $ \Delta g \sim 3$ is used, as a rule.

We use a simple form of the gluon structure function
which contributes to $\sigma$:

$$  g(y)=\frac{R}{y} (1-y)^5,\;\;\;\;R=3. $$

The resulting asymmetry is proportional to the ratio
\be
 C_g= \frac{\Delta g}{R}.
\ee
For  $ \Delta g \sim 3$ we have  $C_g \sim 1$.
This magnitude will be used in what follows. However, it was mentioned in
\ci{rams} that the magnitude  $ \Delta g \sim 1$ is more preferable now.
In this case $C_g$ will be about .3 and the results will decrease by factor 3.

For a standard form of the pomeron vertex (\ref{pmu}) we find \ci{golall}
\be
A_{ll}=\frac{-2 x_p (\ln{\frac{|t|}{M_Q^2}}-3)}{\ln{\frac{|t|}{M_Q^2}}
(2 \ln{\frac{s x_p}{4 |t|}}+\ln{\frac{|t|}{M_Q^2}})}C_g.
\ee
For the pomeron vertex (\ref{ver}) the axial-like term
$V^{\mu}(k,r) \propto  u_3(r) \epsilon^{\mu\alpha\beta\rho}
k_\alpha q_\beta \gamma_\rho \gamma_5$ is extremely  important in
asymmetry. The formula for asymmetry is more complicated in this case.

Our predictions for $A_{ll}$ asymmetry at $\sqrt{s}=40GeV$ (HERA-N energy)
and $x_p=0.2$ for a standard quark-pomeron vertex (\ref{pmu})
and spin-dependent quark-pomeron vertex (\ref{ver}) are shown
in Fig.1 for light quarks and in Fig.2 for a heavy (C) quark.
It is easy to see that the obtained asymmetry strongly depends on the
structure of the quark-pomeron vertex.
For a spin-dependent quark-pomeron vertex $A_{ll}$ asymmetry is smaller by
factor 2 because $\sigma$ in (\ref{asydef}) is larger in this case due to the
contribution of other $u_i$ structures.

As it was mentioned above, the asymmetry in $pp$ polarized diffractive
reactions depends on the unknown the  spin--gluon structure function
$\Delta g$ of the proton.
To obtain more explicit results, let us study the $Q \bar Q$ diffractive
production in a lepton-proton reaction at small $x_p$.

 The standard set of kinematical variables looks as follows \ci{h1}
\ba
s=(p_l+p)^2,\; \;Q^2=-q^2,\;t=(p-p')^2 \nonumber
\\  y=\frac{pq}{p_l p},\;x=\frac{Q^2}{2pq},\;
\beta=\frac{Q^2}{2q(p-p')},\;x_p=\frac{q(p-p')}{qp},
\ea
where $p_l,p'_l$ and $p, p'$ are initial and final lepton and proton
momenta, respectively, $q=p_l-p'_l$.

The asymmetry is determined by formula (\ref{asydef}).
The main contributions to $A_{ll}$ asymmetry in the discussed region
are determined by the $u_0$ and $u_3$ structures in (\ref{ver}).
The formulae
for $\sigma$ and $\Delta \sigma$ for different forms of the pomeron vertex
can be found in \ci{gollp}.
Note that $\Delta \sigma$ is proportional to $Q^2$. As a result, the
asymmetry must increase
with $Q^2$.

Our predictions for $A_{ll}$ asymmetry for the maximal HERA
energy $\sqrt s=300GeV$  estimated
>from perturbative vertex functions for $y=0.5$ and
$x_p=0.2$ for a standard quark--pomeron vertex
and a spin-dependent quark--pomeron vertex are shown in Fig.3.
In this figure one can see the $Q^2$ dependence of $A_{ll}$ for
fixed $|t|=3GeV^2$. The obtained asymmetry is not small and strongly depends
on the spin structure of the quark-pomeron vertex.
Asymmetry decreases with growing $|t|$  and increases with growing $Q^2$.

The single-spin asymmetry differs from the longitudinal asymmetry. It is
 determined by the relation
\be
A_{\perp}=\frac{\sigma(^{\uparrow})-\sigma(^{\downarrow})}
{\sigma(^{\uparrow})+\sigma(^{\downarrow})} =\frac{\Delta \sigma}{\sigma}
\propto
 \frac{\Im (f_{+}^{*} f_{-})}{|f_{+}|^2 +|f_{-}|^2},
\ee
where $f_{+}$ and $f_{-}$ are spin-non-flip and spin-flip amplitudes,
respectively. So, the single--spin asymmetry appears if both $f_{+}$
and $f_{-}$
are nonzero and there is a phase shift between these amplitudes.

For elastic reactions this asymmetry can be predominated by the so--called
"soft pomeron" that includes the rescatterings effects.
For this pomeron the amplitudes $f_{+}$ and $f_{-}$
can possess a phase shift. As a result the transverse hadron asymmetry
determined by the pomeron exchange

\be
A^h_{\perp} \simeq \frac{2m \sqrt{|t|} \Im (AB^{*})}{|B|^2}. \lab{epol}
\ee
appears. Here amplitudes $A$ and $B$ determined in (\ref{prver}) are related
with the proton wave
function and can be calculated by model approaches (see \ci{golpp} e.g.).
The model \ci{golpp} predicts that polarization at $|t| \simeq 1GeV^2$
can be about $10 \div 15\%$.

Let us investigate the single transverse spin asymmetry in
$p\uparrow p \to p+Q \bar Q+X$ reaction.
Standard kinematical variables were determined in (\ref{varpp}). In what
follows we shall calculate the distributions of jets over $p_{\perp}^2$.

The cross sections $\sigma$ and $\Delta \sigma$ can be written in the form
\be
\frac{d \sigma(\Delta \sigma)}{dx_p dt dp_{\perp}^2}=\{1,A^h_{\perp}\}
\frac{\beta^4 |F_p(t)|^2 \alpha_s}{128 \pi s x_p^2}
\int_{4p_{\perp}^2/sx_p}^{1} \frac{dy g(y)}{\sqrt{1-4p_{\perp}^2/syx_p}}
\frac{ N^{\sigma(\Delta \sigma)}
(x_p,p_{\perp}^2,u_i,|t|)}{(p_{\perp}^2+M_Q^2)^2}. \lab{si}
\ee
Here $g$ is the gluon structure function
of the proton, $p_{\perp}$ is a transverse momentum of jets, $M_Q$
is a quark mass, $N^{\sigma(\Delta \sigma)}$ is a trace over the quark loop,
$\beta$ is a pomeron coupling constant, $F_p$ is a pomeron-proton form factor.
In (\ref{si}) the coefficient equal to unity appears in $\sigma$ and
the transverse hadron asymmetry $A^h_{\perp}$ at the pomeron-proton vertex
(\ref{epol}) appears in $\Delta \sigma$.

In the diffractive--jet production investigated here  the main contribution
is determined by the region where the quarks
in the loop are not far of the mass shell.
So, we can assume that the asymmetry factor in (\ref{si}) can be determined
by the soft pomeron and it coincides with the elastic transverse hadron
asymmetry (\ref{epol}). In our further estimations we use magnitude
$A^h_{\perp}=0.1$. New model calculations for the pomeron-proton coupling
(\ref{prver}) are very important. For this purpose, the diquark model
\ci{kroll} should be useful.

 Both $\sigma$ and $\Delta \sigma$ have a similar dependence at small $x_p$
$$ \sigma(\Delta \sigma) \propto \frac{1}{x_p^2} $$
This property of (\ref{si}) allows one to study asymmetry at small
$x_p$ where the pomeron exchange is predominated because of a high energy
in the quark-pomeron system.

In calculations we use the magnitude $\beta=2GeV^{-1}$ \ci{don-la}
and the exponential form of the form factor
$$ |F_p(t)|^2=e^{bt} \;\;\;{\rm with}\;\; b=5GeV^2.$$

Our predictions for  asymmetry $A_{\perp}$
at $\sqrt{s}=40GeV$, $x_p=0.05$ and $|t|=1GeV^2$
for a standard quark-pomeron vertex
(\ref{pmu}) and a spin-dependent quark-pomeron vertex (\ref{ver}) are shown
in Fig.4 for light-quark jets.
It was found that the main contributions to $\sigma$ and $\Delta \sigma$
are determined by the $u_0$ and $u_3$ terms in (\ref{ver}).

It is easy to see that the shape of asymmetry is different for standard and
spin-dependent pomeron vertices. In the first case it is approximately
constant in the second it depends on $p^2_{\perp}$. So, from this
asymmetry the structure of the quark-pomeron vertex can be determined.

We calculate the integrated over $p^2_{\perp}$ of jet  cross sections
$\sigma$ and
$\Delta \sigma$, too. The asymmetry obtained from these integrated cross
sections does not depend practically on the quark-pomeron vertex structure.
It can be written in both the cases in the form
\be
A1=\frac{\int dp^2_{\perp} \Delta  \sigma}{\int dp^2_{ \perp}\sigma} =
0.5\;A^h_{\perp}  \lab{a1}
\ee
As a result, the integrated asymmetry (\ref{a1}) can be used for studying of
the transverse hadron asymmetry $A^h_{\perp}$ at the pomeron-proton vertex.

To summarize, we have presented in this report the perturbative QCD analysis
of longitudinal double  and single transverse spin asymmetries in the
diffractive 2-jet production in $lp$ and $pp$ processes.
The spin-dependent contributions to the quark-pomeron
and hadron-pomeron couplings discussed here modify the calculated spin
asymmetries.
The obtained asymmetries in the lepton-proton and proton-proton processes
have some important properties:
\begin{itemize}
\item
 $A_{ll}$ asymmetry  is opposite in
sign for light and heavy quarks;
\item Asymmetry
for open charm production is sufficiently large;
\item Relavent asymetry in diffractive
$J/\Psi$ production can be large too;
\item Asymmetry decreases with energy only logarithmically, $A_{ll} \sim
1/ln(s x_p/(4|t|)$;
\item Double spin $A_{LL}$ asymmetry is equal to zero at $x_p=0$.
So it is better to study it at $x_p=0.1 \div 0.2$.
Asymmetries strongly depend on the structure the of quark-pomeron
vertex;
\item  The
distribution over jets $p_{\perp}^2$ should be more informative in studying
the pomeron vertex structure.
\end{itemize}

The model prediction shows that the $A_{ll}$ and $A_{\perp}$ asymmetry can be
studied  and the information about the spin structure of the quark-pomeron
vertex can be extracted, for instance, from the future spin experiments at
HERA, HERA-N, RHIC.

It should be emphasized that the spin effects obtained here are
completely determined at fixed momenta transfer by large-distance
contributions in quark (gluon) loops. So, they have a nonperturbative
character. The investigation of spin effects in diffractive reactions is an
important test of the spin sector of QCD at large distances.

This work was supported in part by the Russian Fund of Fundamental Research,
Grant 94-02-04616.

\end{document}